\documentclass[preprint,12pt]{elsarticle}
\biboptions{sort&compress,longnamesfirst,numbers}
\usepackage[english]{babel}
\usepackage{geometry}
\usepackage{lineno}
\usepackage{multirow}
\usepackage{graphicx}
\usepackage{tabularx}
\usepackage{mathtools}
\usepackage{threeparttable}
 \newcolumntype{L}{>{\raggedright\arraybackslash}X}
\usepackage{booktabs}
\usepackage{soul}
\usepackage[export]{adjustbox}
\usepackage{subfig}
\usepackage{amsmath}
\usepackage{amssymb}
\usepackage{multirow}
\usepackage[titletoc]{appendix}

\usepackage{colortbl}
\usepackage{url}

\usepackage{breakurl}
\usepackage[breaklinks]{hyperref}
\definecolor{kugray5}{RGB}{224,224,224}
\makeatletter 
\renewcommand\@biblabel[1]{#1.} 
\makeatother
\usepackage{floatrow}
\usepackage[justification=centering]{caption}

\graphicspath{{Figures/}}

\addto\captionsenglish{}
\usepackage[markup=underlined]{changes}

\definechangesauthor[color=red]{DP}
\definechangesauthor[color=blue]{EB}
\definechangesauthor[color=blue]{MP}
\definechangesauthor[color=orange]{APB}
\usepackage{todonotes}


\journal{Applied Energy}
\makeatletter
\def\ps@pprintTitle{%
 \let\@oddhead\@empty
 \let\@evenhead\@empty
 \def\@oddfoot{}%
 \let\@evenfoot\@oddfoot}
\makeatother
\begin{document}

\begin{frontmatter}
\title{Balancing DSO interests and PV system economics with alternative tariffs}

\author[EPFL]{Alejandro Pena-Bello\corref{cor1}}
\cortext[cor1]{Corresponding author. E-mail: alejandro.penabello@epfl.ch}
\address[EPFL]{Photovoltaics and thin film electronics laboratory (PV-LAB), École Polytechnique Fédérale de Lausanne (EPFL), Institute of Electrical and Microengineering (IEM), Neuchâtel, Switzerland}

\author[EPFL]{Robin Junod}
\author[EPFL]{Christophe Ballif}
\author[EPFL]{Nicolas Wyrsch}

\begin{abstract}
Distributed rooftop photovoltaics (PV) is one of the pillars of the energy transition. However, the massive integration of distributed PV systems challenges the existing grid, with high amounts of PV injection possibly leading to over-voltage and reverse power flow, with line and transformer overloading, among other issues. Moreover, the increase in PV self-consumption and consequently the reduction of imported electricity poses a problem in recovering Transmission System Operators (TSOs) and Distribution System Operators (DSOs) grid costs, that until now have been directly linked to the amount of electricity consumed due to the volumetric nature of traditional tariffs. To investigate whether alternative tariffs could mitigate PV impacts at the distribution level without hampering PV development, we assess five electricity tariffs that could help the DSOs to recover the costs of maintaining the distribution grid. Additionally, we evaluate how such tariffs may affect private investment in storage and their impact on three types of low-voltage networks (i.e., urban, semi-urban, and rural). We found that tariffs with a capacity-based component promote further adoption of PV and storage. At the same time, they allow the DSOs to recover the grid cost without incurring relevant economic differences for the customer. However, all assessed tariffs were found to have a limited role in mitigating PV impacts at the distribution level. 
\end{abstract}
\end{frontmatter}
\renewcommand{\thefootnote}{\alph{footnote}}

\begin{table}[h]
\centering
\scalebox{0.8}{
\begin{tabular}{llll}
PV & Photovoltaic & FiT & Feed-in Tariff\\
SC & Self-consumption & SS & Self-sufficiency\\
FT & Flat tariff & DT & Double tariff\\
CT & Capacity-based tariff & LV & Low-voltage\\
TOU & time of use & MV & Medium-voltage\\
EV & Electric vehicle & HP & Heat pump\\
CAPEX & capital expenditure & OPEX & Operational expenditure \\
TSO & Transmission system operator & DSO & Distribution system operator \\
SIA & Swiss Society of Engineers and Architects & IQR & Interquartile range\\
RegBL & Registre
fédéral des bâtiments et des logements  & MILP & Mixed Integer Linear Programming\\
IRENA & International Renewable Energy Agency\\

\end{tabular}}
\caption*{List of abbreviations.}
\end{table}
\section{Introduction}\label{introduction}
The massive integration of distributed photovoltaics (PV) poses several challenges at the distribution level. For instance, PV injection can lead to over-voltage, line, and transformer overloading, and reverse power flow, among other issues \cite{katiraei2011solar}. This is mainly because the distribution grid infrastructure was dimensioned for consumption only and without upstream flow. 

From the transmission system operator (TSO) and the distribution system operator (DSO) perspective, there is a need to recover the costs of the transmission and distribution grid (i.e., grid costs), to ensure the adequate maintenance of the infrastructure and the reliability and stability of the grid \cite{schittekatte2018future}. Until now small consumers are charged for their share of grid costs based on their energy consumption \cite{nelson2018electricity}. Therefore, the recovery of such costs is being offset by the prosumers due to the increased amount of PV self-consumption and it is expected to be hindered since typical electricity tariffs are mainly volumetric (i.e. the final bill corresponds to the amount of electricity consumed, in \,€/kWh) in many regions including Switzerland \cite{holweger2020mitigating}, UK \cite{ofgem2020ofgem}, and Europe, in general, \cite{irena2019time, ACER2021}. However, using volumetric network charges is driving inefficiencies not previously seen in the electricity industry \cite{simshauser2016distribution, perez2017regulatory}. Although an increase in demand from new devices such as electric vehicles (EVs) and heat pumps (HPs) is anticipated, the increased PV penetration, together with distributed energy storage and energy efficiency measures are expected to eventually reduce the final amount of electricity sold to the final user (in kWh) and volumetric contribution to grid costs.

Existing studies have looked at the effects of volumetric tariffs on prosumer behavior and incentives for residential buildings and households in particular. G\"unther, Schill, and Zerrahn \cite{gunther2021prosumage}, considered the impact of three retail tariffs on prosumer behavior in Germany. The authors explored a flat volumetric tariff, a double tariff, and a real-time pricing tariff and further consider different feed-in tariff (FiTs) rates. The results show a reduction in the optimal PV and battery capacities under a double tariff, leading to reduced self-generation while requiring prosumers to contribute more to fixed power sector costs. Al Arrouqi \textit{et al.} \cite{al2019assessment}, considered flat, double, and dynamic tariffs and analyzed their impact on PV profitability in Australia. They found the dynamic tariff to be more economically beneficial for the utility company in comparison to a flat rate tariff, the latter being the tariff that provides the best internal rate of return for the homeowners. Fett, Fraunholz, and Keles \cite{fett2021diffusion}, looked at the long-term impacts of residential PV and battery diffusion on the electricity market taking into account an electricity market simulation, and different scenarios, including different types of tariffs, comprising the use of volumetric and fixed grid charges, as well as the use of FiTs and dynamic export prices (i.e., depending on the electricity market price), and feed-in limitations. The authors found that restrictive measures (fixed grid charges, and feed-in limitations) lead to substantially smaller photovoltaic and storage systems in the medium term, but this effect is smaller in the long term, mainly due to the lower cost of the PV and storage system and higher retail prices in the future.

Conversely, it has been argued that tariffs with a capacity-based component  (i.e. tariffs that are completely or partially based on the maximum measured power demand during a billing period,  in \,€/kW) will be able to produce a more efficient and more equitable price signal \cite{simshauser2016distribution}. This type of tariff has been widely analyzed in the literature, for instance, Avau, Govaerts, and Delarue \cite{avau2021impact}, compared the impacts of different retail tariffs, including a pure capacity-based tariff that charges the user depending on the maximum annual power capacity measured every hour. They found that prosumers are very sensitive to capacity tariffs and that the price of the capacity tariff is hardly relevant. Schittekatte, Momber, and Meeus \cite{schittekatte2018future}, looked into volumetric, capacity-based, and fixed grid costs tariffs, and detected that capacity-based charges (over) incentivized investment in PV and batteries, which severely distort the investment decisions of consumers. On the contrary, Bloch \textit{et al.} \cite{bloch2019impact}, benchmarked five retail tariffs, including one with a capacity-based component depending on the monthly maximum power exchanged with the grid (for both, import and export), and analyzed the trade-offs between the economic viability of privately owned energy systems and their grid usage intensity. The authors found that the capacity-based tariff did not encourage investment in the battery system, since the cost is dominated by the PV injection power, which can be simply curtailed. Finally, these tariffs were found to create more value for residential batteries when combining different applications \cite{penabello2019optimized}.

While some studies take both perspectives, the private investor and the DSO perspective \cite{gunther2021prosumage, bloch2019impact}, the impact of the tariffs and PV and battery scheduling on the network operation is usually out of the scope of the studies. Only very little research has considered the technical impacts of retail tariffs on the low-voltage (LV) network. Pimm, Cockerill, and Taylor \cite{pimm2018time}, analyzed Time-of-Use (TOU) tariffs' impact on import and export peaks at the low voltage level but only took into account the reduction in maximum demand at the aggregated level. Azuatalam, Chapman, and Verbi{\v{c}} \cite{azuatalam2020probabilistic}, used a probabilistic assessment for two volumetric tariffs and two tariffs with a volumetric and a capacity-based component. The authors concluded that flat tariffs perform better than ToU tariffs for mitigating voltage and alleviating line congestion problems, and the addition of a capacity-based component to such flat tariffs helps to reduce network peaks further. The assessment was done for a single LV network and for predefined PV capacities. Schreck \textit{et al.} \cite{schreck2022importance}, considered a local energy market and determined the impact of energy-based grid tariffs with constant, dependent on the network (urban, semi-urban, and rural), and time-variable cost components and power-based tariffs. The results indicate a high potential of power-based tariffs for peak reduction. However, the authors used pre-defined PV and storage capacities. 

The previously mentioned studies deal with the analysis of different volumetric and/or capacity-based tariffs in LV networks. What is missing, however, is a comprehensive analysis that, considering the Energy Transition context, the DSO perspective, and the prosumer's investment profitability, integrates the effects of different tariffs on the optimum size and operation of PV and battery systems, comparing these effects in different LV networks. 

Against this background, we aim to assess alternative electricity tariffs that help to mitigate the impact of distributed energy resources on the grid, without hampering PV development. This study builds on the related work of Holweger \textit{et al.} \cite{holweger2020mitigating}. We use a MILP optimization of the PV and battery sizing and operation from the consumer's point of view at the building level, following a purely economic rationale. Here we include different types of tariffs and two additional types of LV networks, as well as a new evaluation of the results of the power flow analysis, based on statistical analysis, to assess the following three research questions:

\begin{itemize}
    \item What is the impact of alternative electricity tariffs on the adoption and operation of distributed PV-coupled battery systems?
    \item How does the operation of distributed PV-coupled battery systems under alternative electricity tariffs affect the LV network operation?
    \item How does the impact of the tariffs differ depending on the LV network typology?
\end{itemize}

We contribute to the existing literature by shedding light on the effects of alternative tariffs and alternative tariff structures on the adoption of distributed PV-coupled battery systems, preventing undesired outcomes of tariffs. Moreover, the integration of the three perspectives (i.e. from the national Energy Transition goals, the DSO, and the prosumer), provides a comprehensive view of the outcomes of the study, which would prove to be valuable also for policy-makers and regulators in the path towards a decarbonized economy.

Bearing these questions and contributions in mind, we design five alternative retail tariffs. We use three volumetric tariffs and two tariffs that mix a volumetric and a capacity-based component, developed in collaboration with a Swiss DSO. The volumetric tariffs include two Time-of-Use (TOU) tariffs, and a dynamic volumetric tariff based on historical data at the transformer station to replicate the stress at the transformer level and give a price signal to the consumer to reduce their consumption. Whereas the two tariffs with a capacity-based component have different billing horizons, namely monthly and daily. These five tariffs are compared to two reference tariffs that are currently used, a flat and a double tariff. 

Subsequently, these retail tariffs are tested in three typical Swiss LV networks (urban, semi-urban, and rural), using power flow analysis to  evaluate the power exchanges at the transformer station, as well as line overloading, and voltage violations. We look at the year 2025, using the expected cost of the PV and battery systems for such year \cite{irena_esr2017}, that are assumed to be installed at once for all buildings, one full year of demand data with a 15-minute resolution collected back in 2015 \cite{Flexi1_2015, Flexi2_2017}, that is then adjusted to the transformer present-day load at the transformer level \cite{holweger2021privacy}. The proposed method allows us to provide technical and regulatory recommendations which are important for DSOs, regulators, and policy-makers to design new electricity retail tariffs that help to mitigate the impact of large amounts of distributed energy resources at the LV level without hampering PV development and contribute to recover network costs. With a relatively low density and a small number of residents, the Swiss context allows the derivation of insights into the effects of retail tariffs in residential areas. 

The remaining parts of the paper are structured as follows. In Section \ref{methods}, the input data, as well as the methodology of the paper are introduced. In Section \ref{results}, the results are presented. Section \ref{discussion} is a discussion of the implications of our results; limitations and possibilities for future work are also summarised here. Lastly, a conclusion is formulated.

\section{Materials and methods}\label{methods}
Five electricity tariffs are defined in collaboration with a Swiss DSO, Romande Energie, with a forward-looking perspective based on the expected costs of PV and batteries in the year 2025. These tariffs are tested in three real LV networks (urban, semi-urban, and rural), that are representative of the Romande Energie electric distribution grid network. In the first step, we use an optimization framework to assess the impact of the electricity tariffs on the consumer's decision to install PV and batteries at the building level, as well as its operation, using a purely economic rationale. Then, we run a power flow analysis to evaluate the impact of the electricity tariffs on the LV network. It is worth mentioning that EVs and HPs are out of the scope of this study, and therefore, their associated electricity profile is not considered here.

\subsection{Input data}
\subsubsection{Demand and PV data}
Demand data measurements are not available for the selected LV networks, therefore an estimation of the building demand is needed. Data from three sources are used to estimate the load curves for each building within each zone. First, we recover data from the \textit{Registre fédéral des bâtiments et des logements} (RegBL), such as building date of construction or renovation, building/local categories, building surface, dwellings surface, and the number of floors. Second, we use data from the Swiss Society of Engineers and Architects (SIA) norms, such as electrical consumption per building type and age. Third, smart-meter measurements from the Swiss French-speaking area collected in the framework of the Flexi projects, with 15-min resolution for different building categories (i.e., apartment, house, and not residential) \cite{holweger2021privacy, Flexi1_2015, Flexi2_2017} are used as a proxy variable. Finally, we utilize the real load curve at the transformer station provided by Romande Energie to match the sum of individual building loads at the aggregated level.

Additionally, for the three LV networks, information on the roof surface area, azimuth, and tilt for every building inside the selected LV network is extracted using the RegBL. To model the PV generation, monitored environmental variables including outdoor temperature and horizontal solar irradiance of the MeteoSwiss weather station situated in Pully, Vaud are used. These data are combined with data from the Sandia PV module database for the SunPower SPR-315E-WHT PV modules, to calculate the current and voltage performance per module, which are then aggregated depending on the roof orientation. 

Finally, The PV and battery price levels have been set according to the reference year 2025. The price projections have been extracted from the IRENA report \cite{irena_esr2017} and calibrated for the Swiss price levels provided in a recent market study \cite{suisseenergie2021} using the empirical model presented by Bloch \textit{et al.} \cite{bloch2019impact}. The resulting PV and battery cost function is decomposed into a fixed (e.g. administrative procedures and fixed installation costs) and a linear component (relative to the installation capacity) \cite{suisseenergie2021}. These linear cost functions are annualized and used along with the PV yearly maintenance cost and battery lifetime in Eq. \ref{eq:obj}. Table \ref{tab:param} displays the main techno-economic parameters of this study.

\begin{table}[]
\begin{tabular}{lll}
\toprule
Parameter              & Value  &    Reference  \\\midrule
System lifetime        & 25 years  &  \cite{irena_esr2017} \\
Discount rate          & 3\%        & Own assumption\\
PV fixed cost          & 10 049 \,€ &  \cite{irena_esr2017,suisseenergie2021}\\
PV specific costs      & 1.05 \,€/W &  \cite{irena_esr2017,suisseenergie2021}\\
Battery fixed cost     & 0 \,€      &  \cite{irena_esr2017,suisseenergie2021}\\
Battery specific costs & 229 \,€/kWh & \cite{irena_esr2017,suisseenergie2021}\\\bottomrule
\end{tabular}
\caption{Techno-economic parameters used in this study.}
\label{tab:param}

\end{table}

\subsubsection{Low-voltage network data}
\begin{figure}[]
    
\newcommand{\fhe}{4.2cm}
    \centering

    \subfloat[\label{fig:grids_morges} Urban LV network.]{\includegraphics[width=0.5\textwidth]{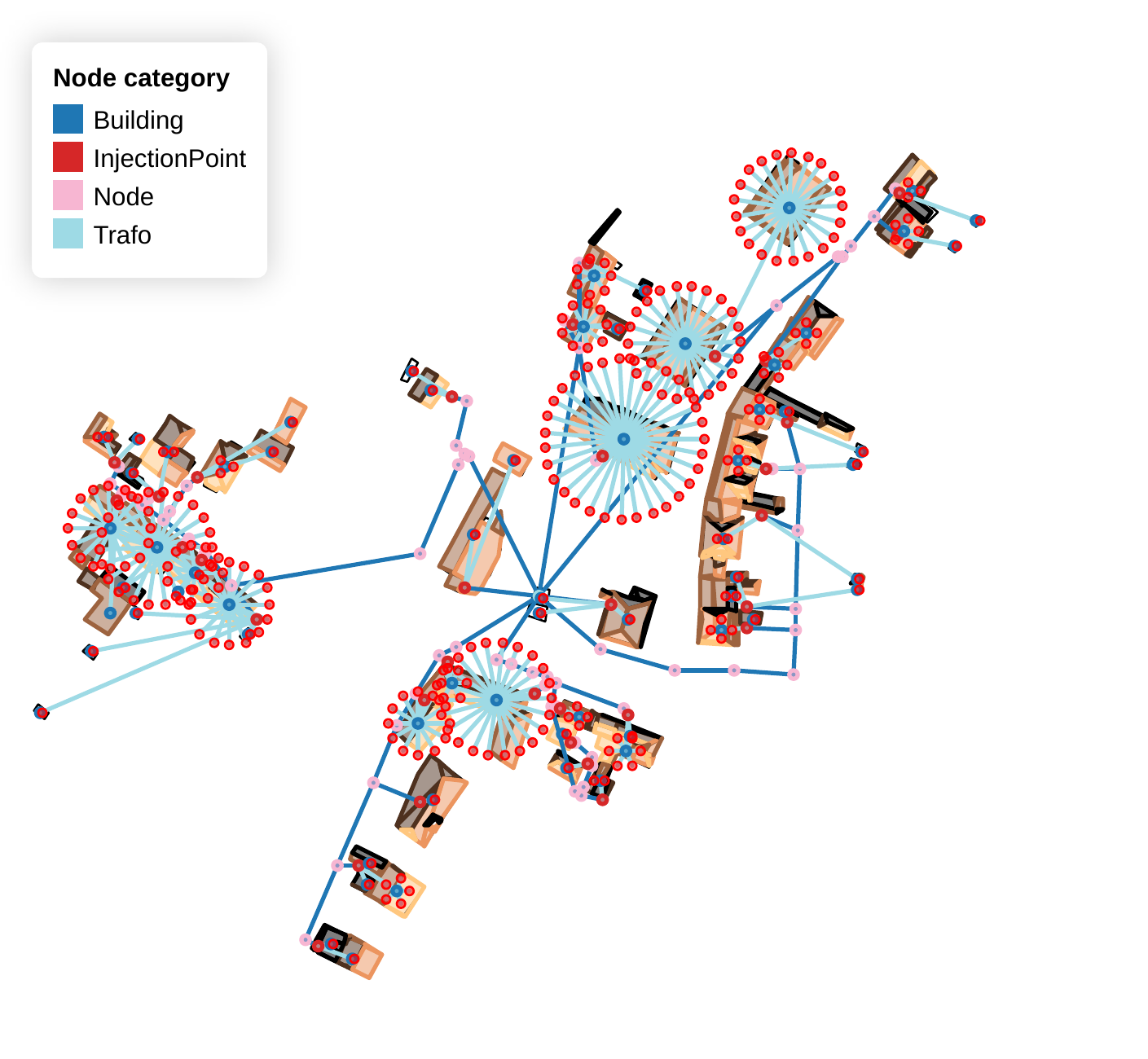}} 
    \subfloat[\label{fig:grids_rolle}Semi-urban LV network.]{ \includegraphics[width=0.5\textwidth]{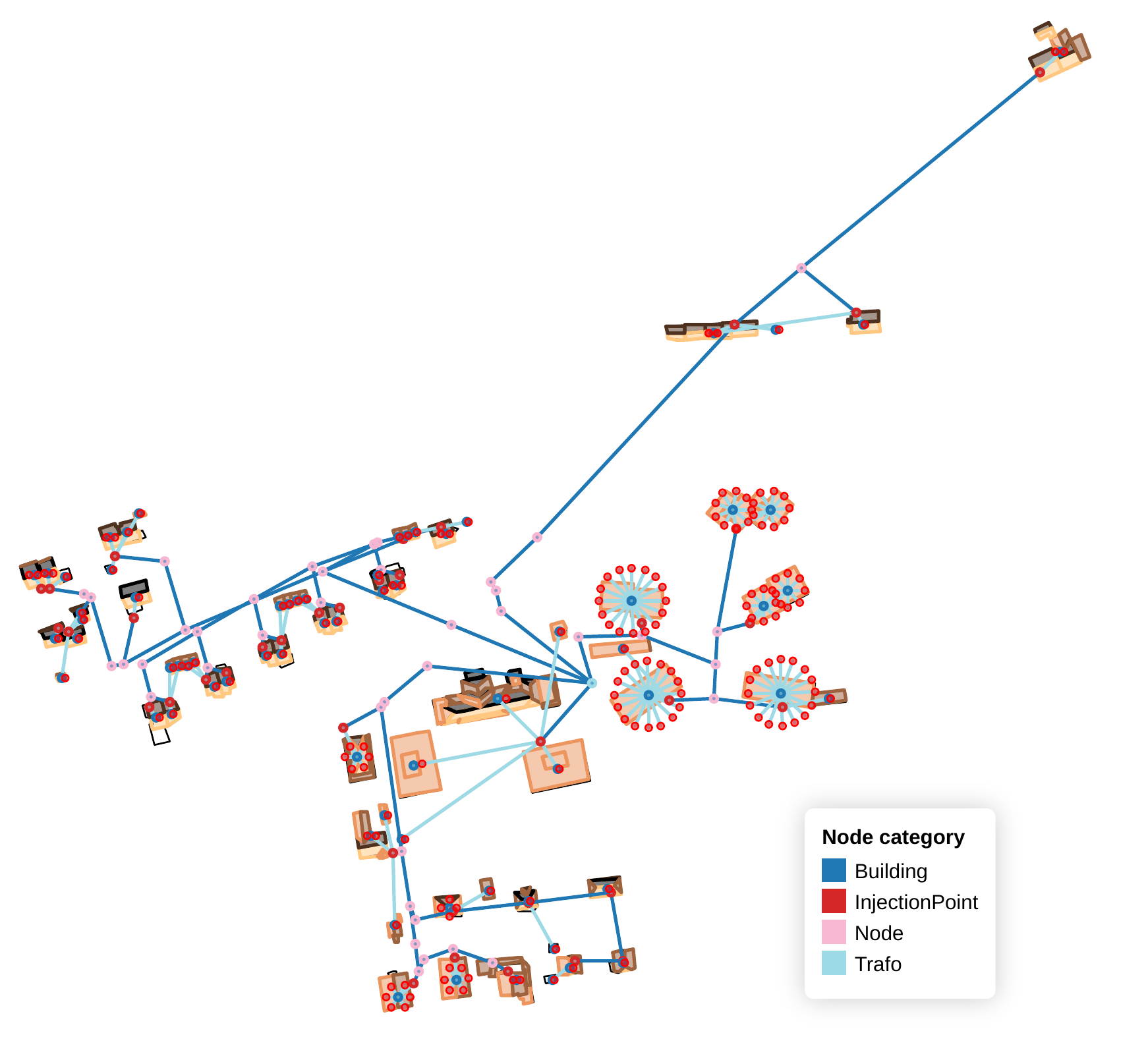}}\\
    
    \subfloat[\label{fig:grids_Essertines} Rural LV network.]{\includegraphics[width=0.5\textwidth]{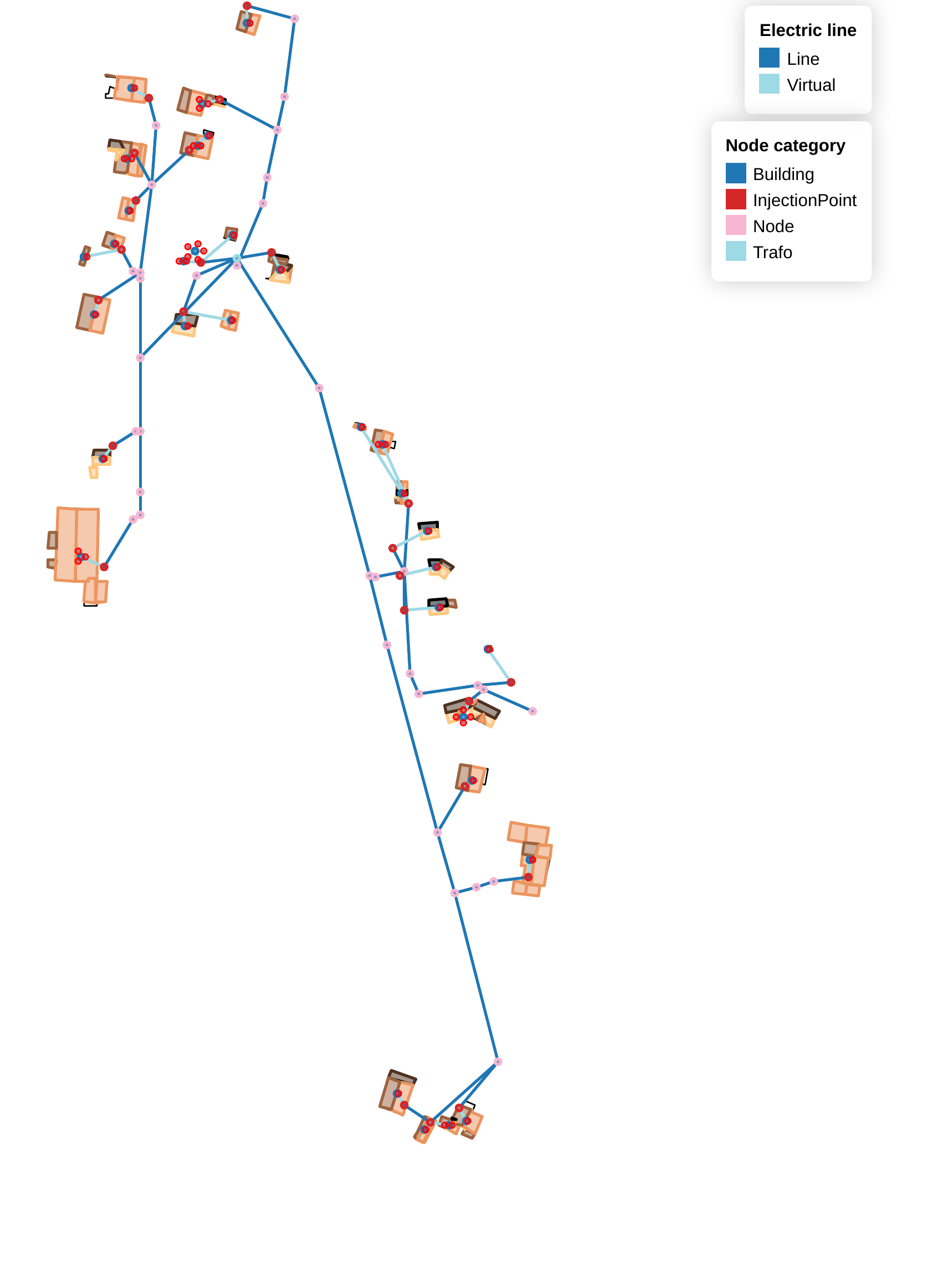}}
    \caption{Illustration of the LV networks used in this study.}
    \label{fig:grids}
\end{figure}

Romande Energie provided for each LV network the transformer, cables, and line characteristics (e.g. capacity, line length, impedance...), as well as the federal building identifier (EGID) of each customer linked to each injection point and the measured load charge of the transformer station, including a correction for the existing PV capacity. The three LV networks were chosen to represent rural, semi-urban, and urban LV networks (see Figure \ref{fig:grids}). Table \ref{tab:grids} presents the main characteristics of each LV network.

\begin{table}[]
\scalebox{0.65}{
\begin{tabular}{llllllllll}
\toprule
LV network type                   & \multicolumn{1}{{m{2.cm}}}{Transformer capacity [kVA]} & \multicolumn{1}{m{2.5cm}}{Virtual Transformer Capacity [kVA]} & 
\multicolumn{1}{{m{2.cm}}}{Number of loads} & 
\multicolumn{1}{{m{2.3cm}}}{Number of buildings} & 
\multicolumn{1}{{m{2.cm}}}{Number of injection points} & 
\multicolumn{1}{{m{2.cm}}}{Max power [kW]} & 
\multicolumn{1}{{m{2.5cm}}}{PV correction max [kW]} & 
\multicolumn{1}{{m{2.5cm}}}{Total consumption [MWh]} \\\midrule
Urban                 & 2x 630                     & 1000                               & 287              & 65                   & 33                          & 419.8               &               -         & 1671.2               \\
Semiurban                  & 2x 400                     & 1000                               & 174              & 71                   & 37                          & 508.6              &           -             & 2065.6             \\
Rural & 1x 630                     & 630                                & 48               & 32                   & 24                          & 75.9                & 59.7                   & 204.3 \\\bottomrule
\end{tabular}}
\caption{Main characteristics of the low-voltage networks.}
\label{tab:grids}

\end{table}

\subsection{Tariffs}\label{tar}
Five alternative tariffs aiming to limit the impact of PV at the distribution level are assessed. We aim to avoid hampering the economic attractiveness of potential rooftop PV installations in order to reach the national PV targets. Moreover, the five tariffs are assessed and compared to two references from Romande Energie (flat and double tariff) \cite{romandeEnergie2022}. These alternative tariffs have two different structures, three are purely volumetric tariffs (i.e. the final bill corresponds to the amount of electricity consumed, in \,€/kWh), and two are mixed tariffs with a volumetric and a capacity-based part (i.e. the tariffs are partially based on the maximum measured power demand during a billing period, in \,€/kW):
\begin{enumerate}
    \item A volumetric DT that seeks to promote consumption when PV production is substantial (i.e., DT solar) with an off-peak time during half of the year (i.e., from April to September) every day before 18 h, and during winter the tariff is the same as the reference DT.
    \item A volumetric double tariff during winter, equal to the reference DT, with a flat tariff during summer (i.e., DT summer flat). This tariff seeks to promote lower consumption during peak hours in winter and to encourage self-consumption in summer.
    \item A purely volumetric dynamic tariff that is proportional to the aggregated expected load curve of Romande Energie for 2025 including projections for PV integration, with peaks at the same time when demand peaks at the aggregated level (for instance, the dynamic tariff is higher in the afternoon than the rest of the day, as well as it is higher in winter than in summer, see Figure A1 in Supplementary Material).
    \item A tariff with a capacity component with a monthly billing horizon (i.e., \,€/$kW_{month-i}$). With this tariff, the idea is to limit the \textbf{maximum monthly power drawn} from the grid and limit the impact of demand on the grid.
    \item A tariff with a capacity component with a daily billing horizon (i.e., €/$kW_{day-j}$). As in the case of the monthly tariff, the rationale is to limit the \textbf{daily maximum power drawn} from the grid and limit the impact of demand on the grid.
\end{enumerate}

The tariffs with the capacity component are a mix of volumetric and capacity-based tariffs, accounting for the volumetric nature of energy procurement and taxes, whereas the capacity component aims to replace the grid cost (see Table \ref{tab:tariffs}). These distribution and transmission costs are designed to match the average of the distribution and transmission costs of the two reference tariffs. For all the tariffs, the taxes (including federal and cantonal duties as well as feed-in remunerations) are maintained volumetric and constant and account for 2.92 \textcent/kWh (please note that taxes are maintained constant across different municipalities for comparability reasons). Finally, the export tariff (i.e., the mandatory take-back/feed-in tariff) is kept constant across tariffs at 9.5 \textcent/kWh, whereas the fixed part of the tariff (subscription, in \,€/month) is set to zero in order to do a fair comparison among the seven tariffs. Table \ref{tab:tariffs} presents the main components of each tariff.


A calibration of these tariffs was performed aiming at keeping the DSO revenue (from selling energy minus the cost of buying exported power) of each network as close as possible to the net profit of the reference scenario, under the hypothesis that all buildings will behave exactly the same way as under the reference scenario. This process was done empirically through the adjustment of the energy costs and the grid costs in the case of purely volumetric tariffs. Whereas in the case of tariffs with a capacity component, the adjustment was done on the energy costs and the import power cost (see Table \ref{tab:tariffs}).

Please note that the taxes and the export tariff are kept constant across the different tariffs. In the case of the dynamic tariff, the energy cost and the grid cost are considered to be equal (i.e., 50/50 distribution after the subtraction of taxes). Finally, the average of the results for the three grid networks was used for the final results presented in this report, to avoid local pricing and enable fair comparisons between the networks.

\begin{table}[]
\scalebox{0.5}{
\begin{threeparttable}

\begin{tabular}{lllllllllll}
\toprule
\multicolumn{1}{c}{Tariff}                                   & \multicolumn{1}{c}{Season\tnote{a}}                  & \multicolumn{1}{c}{Peak}           & \multicolumn{1}{{m{2.cm}}}{Energy cost {[}\textcent/kWh{]}} & \multicolumn{1}{{m{1.8cm}}}{Grid cost  {[}\textcent/kWh{]}} & \multicolumn{1}{{m{2.5cm}}}{Tax {[}\textcent/kWh{]}} & \multicolumn{1}{{m{2cm}}}{Total cost {[}\textcent kWh{]}} & \multicolumn{1}{{m{2cm}}}{FiT   {[}\textcent/kWh{]}} & \multicolumn{1}{{m{2.5cm}}}{Fixed Fee   {[}\,€/month{]}} & \multicolumn{1}{{m{2.5cm}}}{Import power cost  {[}\,€/kW{]}} & \multicolumn{1}{{m{2.5cm}}}{Export   power cost {[}\,€/kW{]}} \\\midrule

\multicolumn{1}{c}{\textbf{Reference   FT}}                           & \multicolumn{1}{c}{-}                       & \multicolumn{1}{c}{-}              & \multicolumn{1}{c}{7.98}                      & \multicolumn{1}{c}{8.45}                 & \multicolumn{1}{c}{2.92}                & \multicolumn{1}{c}{19.35}                    & \multicolumn{1}{c}{9.5}                 & \multicolumn{1}{c}{0}                           & \multicolumn{1}{c}{-}                               & \multicolumn{1}{c}{-}                                \\\midrule

\multicolumn{1}{c}{\multirow{2}{*}{\textbf{Reference DT}}}            & \multicolumn{1}{c}{\multirow{2}{*}{-}}      & \multicolumn{1}{c}{Peak   (M-F 6-22)}  & \multicolumn{1}{c}{9.17}                      & \multicolumn{1}{c}{9.86}                 & \multicolumn{1}{c}{2.92}                & \multicolumn{1}{c}{21.95}                    & \multicolumn{1}{c}{9.5}                 & \multicolumn{1}{c}{0}                           & \multicolumn{1}{c}{-}                               & \multicolumn{1}{c}{-}                                \\
\multicolumn{1}{c}{}                                         & \multicolumn{1}{c}{}                        & \multicolumn{1}{c}{Off-Peak}       & \multicolumn{1}{c}{5.87}                      & \multicolumn{1}{c}{5.26}                 & \multicolumn{1}{c}{2.92}                & \multicolumn{1}{c}{14.05}                    & \multicolumn{1}{c}{9.5}                 & \multicolumn{1}{c}{0}                           & \multicolumn{1}{c}{-}                               & \multicolumn{1}{c}{-}                                \\\midrule
\multicolumn{1}{c}{\multirow{4}{*}{\textbf{DT solar}}}                & \multicolumn{1}{c}{\multirow{2}{*}{Summer}} & \multicolumn{1}{c}{Peak   (18-00)\tnote{b}} & \multicolumn{1}{c}{9.17}                      & \multicolumn{1}{c}{9.86}                 & \multicolumn{1}{c}{2.92}                & \multicolumn{1}{c}{21.95}                    & \multicolumn{1}{c}{9.5}                 & \multicolumn{1}{c}{0}                           & \multicolumn{1}{c}{-}                               & \multicolumn{1}{c}{-}                                \\
\multicolumn{1}{c}{}                                         & \multicolumn{1}{c}{}                        & \multicolumn{1}{c}{Off-Peak}       & \multicolumn{1}{c}{5.87}                      & \multicolumn{1}{c}{5.26}                 & \multicolumn{1}{c}{2.92}                & \multicolumn{1}{c}{14.05}                    & \multicolumn{1}{c}{9.5}                 & \multicolumn{1}{c}{0}                           & \multicolumn{1}{c}{-}                               & \multicolumn{1}{c}{-}                                \\
\multicolumn{1}{c}{}                                         & \multicolumn{1}{c}{\multirow{2}{*}{Winter}} & \multicolumn{1}{c}{Peak (M-F 6-22)}           & \multicolumn{1}{c}{9.17}                      & \multicolumn{1}{c}{9.86}                 & \multicolumn{1}{c}{2.92}                & \multicolumn{1}{c}{21.95}                    & \multicolumn{1}{c}{9.5}                 & \multicolumn{1}{c}{0}                           & \multicolumn{1}{c}{-}                               & \multicolumn{1}{c}{-}                                \\
\multicolumn{1}{c}{}                                         & \multicolumn{1}{c}{}                        & \multicolumn{1}{c}{Off-Peak}       & \multicolumn{1}{c}{5.87}                      & \multicolumn{1}{c}{5.26}                 & \multicolumn{1}{c}{2.92}                & \multicolumn{1}{c}{14.05}                    & \multicolumn{1}{c}{9.5}                 & \multicolumn{1}{c}{0}                           & \multicolumn{1}{c}{-}                               & \multicolumn{1}{c}{-}                                \\\midrule
\multicolumn{1}{c}{\multirow{3}{*}{\textbf{DT summer flat}}} & \multicolumn{1}{c}{Summer}                  & \multicolumn{1}{c}{FT}             & \multicolumn{1}{c}{7.98}                      & \multicolumn{1}{c}{8.45}                 & \multicolumn{1}{c}{2.92}                & \multicolumn{1}{c}{19.35}                    & \multicolumn{1}{c}{9.5}                 & \multicolumn{1}{c}{0}                           & \multicolumn{1}{c}{-}                               & \multicolumn{1}{c}{-}                                \\
\multicolumn{1}{c}{}                                         & \multicolumn{1}{c}{\multirow{2}{*}{Winter}} & \multicolumn{1}{c}{Peak (M-F 6-22)}    & \multicolumn{1}{c}{9.17}                      & \multicolumn{1}{c}{9.86}                 & \multicolumn{1}{c}{2.92}                & \multicolumn{1}{c}{21.95}                    & \multicolumn{1}{c}{9.5}                 & \multicolumn{1}{c}{0}                           & \multicolumn{1}{c}{-}                               & \multicolumn{1}{c}{-}                                \\
\multicolumn{1}{c}{}                                         & \multicolumn{1}{c}{}                        & \multicolumn{1}{c}{Off-Peak}       & \multicolumn{1}{c}{5.87}                      & \multicolumn{1}{c}{5.26}                 & \multicolumn{1}{c}{2.92}                & \multicolumn{1}{c}{14.05}                    & \multicolumn{1}{c}{9.5}                 & \multicolumn{1}{c}{0}                           & \multicolumn{1}{c}{-}                               & \multicolumn{1}{c}{-}                                \\\midrule
\multicolumn{1}{c}{\textbf{Dynamic}}                    & \multicolumn{1}{c}{-}                 & \multicolumn{1}{c}{-} & \multicolumn{1}{c}{10.76\tnote{c}} & \multicolumn{1}{c}{10.76\tnote{c}} & \multicolumn{1}{c}{2.92} & \multicolumn{1}{c}{24.44\tnote{c}} & \multicolumn{1}{c}{9.5} & \multicolumn{1}{c}{0} & \multicolumn{1}{c}{-} & \multicolumn{1}{c}{-} \\\midrule

\multicolumn{1}{c}{\textbf{CT   Monthly}}              & \multicolumn{1}{c}{-}                       & \multicolumn{1}{c}{-}              & \multicolumn{1}{c}{7.98}                      & \multicolumn{1}{c}{-}                    & \multicolumn{1}{c}{2.92}                & \multicolumn{1}{c}{10.9}                     & \multicolumn{1}{c}{9.5}                 & \multicolumn{1}{c}{0}                           & \multicolumn{1}{c}{16.4}                         & \multicolumn{1}{c}{0}                                \\\midrule
\multicolumn{1}{c}{\multirow{3}{*}{\textbf{CT Daily}}} & \multicolumn{1}{c}{Summer}                  & \multicolumn{1}{c}{-}               & \multicolumn{1}{c}{7.98}                      & \multicolumn{1}{c}{-}                    & \multicolumn{1}{c}{2.92}                & \multicolumn{1}{c}{10.9}                     & \multicolumn{1}{c}{9.5}                 & \multicolumn{1}{c}{0}                           & \multicolumn{1}{c}{0.5312}                          & \multicolumn{1}{c}{0}                                \\
\multicolumn{1}{c}{}                                         & \multicolumn{1}{c}{Fall/Spring}                   & \multicolumn{1}{c}{-}               & \multicolumn{1}{c}{7.98}                      & \multicolumn{1}{c}{-}                    & \multicolumn{1}{c}{2.92}                & \multicolumn{1}{c}{10.9}                     & \multicolumn{1}{c}{9.5}                 & \multicolumn{1}{c}{0}                           & \multicolumn{1}{c}{0.9296}                          & \multicolumn{1}{c}{0}                                \\
\multicolumn{1}{c}{}                                         & \multicolumn{1}{c}{Winter}                  & \multicolumn{1}{c}{-}              & \multicolumn{1}{c}{7.98}                      & \multicolumn{1}{c}{-}                    & \multicolumn{1}{c}{2.92}                & \multicolumn{1}{c}{10.9}                     & \multicolumn{1}{c}{9.5}                 & \multicolumn{1}{c}{0}                           & \multicolumn{1}{c}{1.3280}                          & \multicolumn{1}{c}{0}                                \\\bottomrule
\end{tabular} 
\begin{tablenotes}
\normalsize{\item[a] When the year is divided into two seasons. The summer runs from the $1^{st}$ April to the $30^{th}$ September.}
\normalsize{\item[b] The on-peak tariff applies every day of the week.}
\normalsize{\item[c] The values of the dynamic tariff are variable throughout  the year, here we present the median values for the energy cost and the grid cost, whereas the taxes are constant throughout the year.}

\end{tablenotes}
\end{threeparttable}}
\caption{Electricity tariff components depending on the tariff used in this study. See Section \ref{tar} for details on the calibration process followed to obtain the coefficients presented in this table.}
\label{tab:tariffs}
\end{table}
\subsection{Modeling}
The modeling is done in three stages. First, the demand allocation for each LV network, followed by the optimization from the consumer perspective, based on a purely economic rationale, and finally, the power flow analysis for each LV network is carried out.
\subsubsection{Demand allocation}\label{sec:dem}
The methodology for demand allocation was developed by Holweger \textit{et al.}\cite{holweger2021privacy}, and it relies on a two-stage Mixed-integer linear programming (MILP) optimization. The allocation process is separated into three steps. First, the annual energy consumption of each building within each predefined zone is estimated using the RegBL data and SIA norms, describing the load category of each building (e.g., apartment, house, not-residential) \cite{holweger2021privacy, Flexi1_2015, Flexi2_2017}. Then, the first stage of the optimization is to allocate and scale a load curve to each building depending on its category and annual consumption. For this, a minimization of the difference between the annual energy consumption for each building and the reference load curves from real smart-meter data is carried out. Finally, the second stage of the optimization minimizes the difference between the sum of all building load curves and the measured transformer load curve, while the changes done to every building load curve are also minimized (see Supplementary Information Section 3 for the model validation). 

\subsubsection{Design and battery scheduling optimization}
Building upon previous work \cite{bloch2019impact, holweger2020mitigating}, the PV and battery sizing and operation for each building are optimized, minimizing the total cost of ownership, i.e., the sum of the annualized investment, maintenance, and operational cost (see Eq. \ref{eq:obj}), subject to power balance constraints (see Refs. \cite{bloch2019impact, holweger2020mitigating}, Supplementary Information Section 2 for further information on modeling, and Supplementary Information Section 3 for the model validation). In this work, battery charging from the grid and battery export are forbidden, according to existing Swiss regulations for behind-the-meter storage systems. Therefore, the battery can uniquely be used to charge from the PV system and to cover the building's electricity needs. Perfect forecasting is used for both PV generation and electrical demand. 

\begin{equation}
\label{eq:obj}
minimize ~ (\textsc{capex}(P_{CAP}^{PV},E_{CAP}^{BAT}) + \textsc{opex}(P_{t}^{IMP},P_{t}^{EXP}))
\end{equation}

where \textsc{capex} is a function of the installed PV capacity ($P_{CAP}^{PV}$) and the battery size $(E_{CAP}^{BAT})$, \textsc{opex} are a function of the grid exchange power ($P_{t}^{IMP}$,$P_{t}^{EXP}$). The objective function is subject to battery constraints and power balance constraints (for the load, the grid, and the PV system). The operational costs are the cost of exchanging energy with the grid (maintenance costs are not taken into account). We consider volumetric and capacity-based tariffs (Eqs. \ref{eq:ge_vol} and \ref{eq:ge_cap}).

\begin{equation}
    \label{eq:OPEX}
     \textsc{OPEX} = \textsc{ox}_\text{ge}^\text{vol} + \textsc{ox}_\text{ge}^\text{pow}\\
\end{equation}
\begin{subequations}
	\begin{align}
	\label{eq:ge_vol}
	& \text{Volumetric tariff} & \textsc{ox}_\text{ge}^\text{vol} &= \sum_{t=1}^T \left[ P^\textsc{imp}_t \cdot  t^\textsc{imp}_t  - P^\textsc{exp}_t \cdot  t^\textsc{exp}_t\right ]\cdot \textsc{ts}_t\\
	\label{eq:ge_cap}
	& \text{Capacity-based tariff} & \textsc{ox}_\text{ge}^\text{pow} &= \sum_{k=1}^K  P^\textsc{max}_{k} \cdot  t^\textsc{max}_k
\end{align}
\end{subequations}

where $t^\textsc{imp}_t$ and $t^\textsc{exp}_t$ are the volumetric import and export tariff respectively
(in \,€/kWh), and $\textsc{ts}_t$ is the simulation time-step. For the capacity-based tariff, the maximum power of the billing period k, $P^\textsc{max}_k$ and $t^\textsc{max}_k$, is the maximum import power per billing period k. Two of the tariffs assessed in this study include a capacity-based component (namely tariffs CT monthly and CT daily), but they also count with a volumetric component, therefore, the \textsc{opex} is the sum of both, volumetric and capacity-based tariff components (see Eq. \ref{eq:OPEX}).

\subsubsection{Power flow analysis}
Taking as input the results of the PV-coupled battery system scheduling optimization, and the LV network characteristics, a graph of the grid network is defined using python packages Networkx and \textsc{pandapower}. We use the per-unit (p.u.) system for voltage at every network node and for current flowing through every line. 

Line loading level is used as a metric for grid congestion, given the lines’ properties, in particular, the maximum allowable current. The line loading level is the ratio between the current and the maximum nominal current of the line; we consider an overloaded line when the line loading level is higher than 100\%. Moreover, we consider the situation separately when the bus voltage at an injection point is above 1\,p.u. and when it is below 1\,p.u., and use the $95^{th}$ percentile of the bus voltage deviation, to distinguish when there is a local excess of energy from when there is a local deficit of energy. We consider voltage violations to be higher than 1.1\,p.u. and lower than 0.9\,p.u. \cite{hartvigsson2021estimating}. 

Finally, we use load duration curves to assess the violations of the transformer power capacity for reverse power flow from LV towards upstream medium-voltage (MV) networks, as well as the transformer permissible overload for different periods of time.

\subsection{Key performance indicators}
We use four technical indicators to evaluate the five considered tariffs' performance. First, PV self-consumption,  which is the share of on-site PV generation that is used to cover the local electricity demand, and self-sufficiency (SS), which is the share of local demand that is covered by the on-site PV generation as shown in Eqs. \ref{eq:TSC} and \ref{eq:SS}. Moreover, we present the results of the total amount of PV and storage installed per grid network.

\begin{equation}
SC={\frac{\sum_{t}{(P^{\textsc{pv}-\textsc{load}}_t+P^{\textsc{pv}-\textsc{batt}}_t)}}{\sum_{t}{P^{\textsc{pv}}_t}}}
\label{eq:TSC}
\end{equation}
\begin{equation}
SS={\frac{\sum_{t}{(P^{\textsc{pv}-\textsc{load}}_t+P^{\textsc{batt}-\textsc{load}}_t)}}{\sum_{t}{P^\textsc{load}_t}}}
\label{eq:SS}
\end{equation}

where $P^{\textsc{pv}-\textsc{load}}_t$ is the share of PV generation that directly meets local electricity demand;  $P^{\textsc{pv}-\textsc{batt}}_t$ is the share of PV generation that is charged into the battery; $P^{\textsc{PV}}_t$ is the total PV generation; $P^{\textsc{batt}-\textsc{load}}_t$ is the amount of electricity discharged from the battery to cover local electricity demand ($P^\textsc{load}_t$).

We analyze the PV hosting capacity of the LV network, i.e. the maximum amount of PV that can be accommodated in a power grid within current regulations without requiring reinforcement, taking into account not only the transformer capacity but also the permissible overload for different timespans. Finally, we calculate two economic indicators, the total cost of the bill per building and aggregated per zone, and the portion corresponding to the sum of distribution and transmission costs, to evaluate the grid cost recovery. 

\section{Results}\label{results}
The results are presented in three sections: Firstly, the impact of tariffs on investment decisions (i.e. PV and battery investment) and energy imports and exports. Secondly, the impact on the grid, covering the load duration curve at the transformer as well as the permissible overload for varying periods of time, voltage deviation and line loading levels and grid usage, and finally, the impact on the grid cost recovery is examined.

\subsection{Tariff's impact on investment decisions and energy consumption}
We compare the electricity tariffs first in terms of PV and storage capacity installed, as well as differences in energy import and export, and secondly in terms of self-consumption and self-sufficiency.

Table \ref{tab:res} presents the differences per electricity tariff for PV and battery capacity installed, as well as yearly energy imports and exports as a result of the optimization from the building perspective. In terms of PV capacity, the tariffs with a capacity-based component (i.e. CT daily and CT monthly) allow more PV installations to be operated within the three studied zones (1 to 5 additional percentage points more than in the reference cases). However, the differences in total PV capacity are smaller than 5\% and most of the potential is exploited. All simple and double tariffs reach similar results, with very similar PV and storage capacity installed. As a response to variable prices in the case of dynamic tariffs and to the capacity component of CT daily and CT monthly tariffs, the rational prosumer will install energy storage. As a result of battery installations, energy imports decline up to 64\%, 39\%, and 37\% for  rural, semi-urban, and urban, respectively, for the CT monthly tariff, compared to the FT reference. 

Conversely, energy exports are slightly affected in the rural zone, with a reduction of 2\%, when comparing CT monthly with the FT reference. In the cases of semi-urban and urban zones, the maximum differences are 16\%, and 28\%, respectively. Despite the relatively small difference in installed PV capacity in the reference FT and CT monthly cases (i.e., an additional 52, 12, and 66\,kW for the rural, semi-urban, and urban zones, respectively), the increase in the aggregated annual yield of the PV systems (i.e., equivalent to 970\,kWh/kW), together with the increase in self-consumption due to the differences in the installed battery capacity, account for the differences in annual import and export energy. At the building level, among all volumetric tariffs, the differences in the amount of energy imported are not statistically significant (p-values$>$0.1 when compared to the two references). Energy export differences are not statistically significant for either type of tariff (p-values$>$0.1 when compared to the two references). 

\begin{table}[ht]
\centering
\scalebox{0.65}{
\begin{tabular}{clccccc}
\hline
\multirow{2}{*}{Network} & \multirow{2}{*}{Tariff} & PV Capacity & PV penetration & Battery Capacity & Energy Import & Energy Export \\ 
  &  & kW & \% & kWh & MWh p.a. & MWh p.a. \\ 
  \hline
\multirow{6}{*}{\rotatebox[origin=c]{90}{Rural}}&    FT reference & 1085 & 94 & 1 & 108 & 1006 \\ 
   & DT reference & 1085 & 94 & 1 & 107 & 1006 \\ 
   & DT summer flat & 1085 & 94 & 1 & 107 & 1006 \\ 
   & DT solar & 1084 & 94 & 1 & 108 & 1005 \\ 
   & Dynamic & 1111 & 96 & 112 & 76 & 999 \\ 
   & CT daily & 1125 & 98 & 229 & 55 & 992 \\ 
   & CT monthly & 1137 & 99 & 399 & 39 & 987 \\  
   \hline
\multirow{5}{*}{\rotatebox[origin=c]{90}{Semi-urban}}& FT reference & 1384 & 92 & 6 & 520 & 1274 \\ 
   & DT reference & 1384 & 92 & 15 & 517 & 1271 \\ 
   & DT summer flat & 1384 & 92 & 10 & 519 & 1273 \\ 
   & DT solar & 1360 & 91 & 16 & 521 & 1248 \\ 
   & Dynamic & 1384 & 92 & 396 & 424 & 1173 \\ 
   & CT daily & 1396 & 93 & 888 & 341 & 1097 \\ 
   & CT monthly & 1396 & 93 & 1160 & 316 & 1072 \\ 
   \hline
\multirow{6}{*}{\rotatebox[origin=c]{90}{Urban}}   & FT reference & 1711 & 95 & 69 & 1113 & 1295 \\ 
   & DT reference & 1711 & 95 & 79 & 1110 & 1291 \\ 
   & DT summer flat & 1711 & 95 & 83 & 1107 & 1289 \\ 
   & DT solar & 1703 & 95 & 52 & 1121 & 1298 \\ 
   & Dynamic & 1741 & 97 & 621 & 967 & 1172 \\ 
   & CT daily & 1777 & 99 & 2092 & 754 & 985 \\ 
   & CT monthly & 1777 & 99 & 3338 & 701 & 930 \\    
   \hline
\end{tabular}
}
\caption{Cumulative installed capacities and energy imports and exports per low-voltage network and per tariff.}
\label{tab:res}
\end{table}

\begin{figure}
\centering
\includegraphics[scale=0.5]{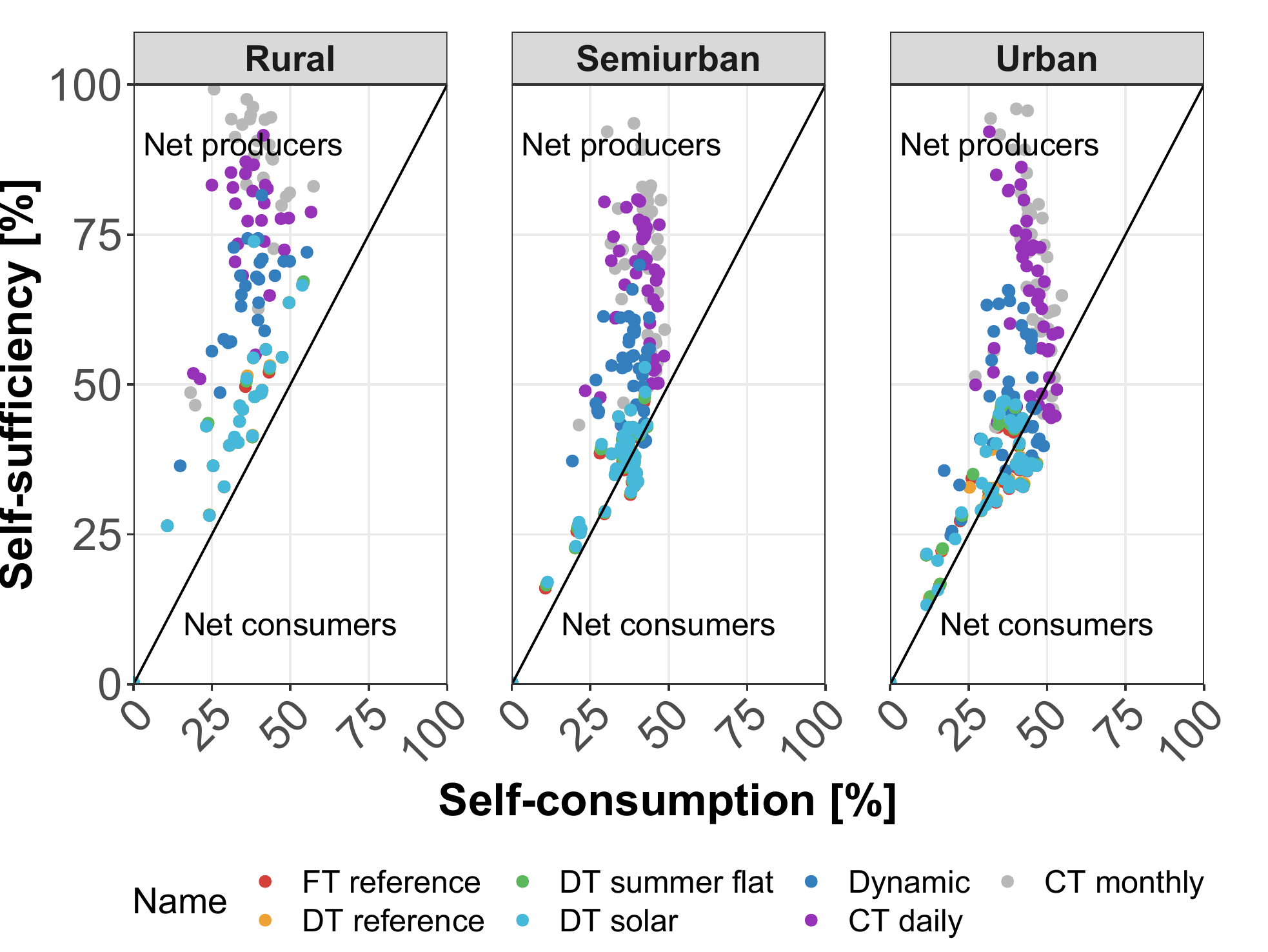}{\centering}
\caption{Energy matching chart to analyze self-consumption (SC) and self-sufficiency (SS) per network. Please note that each point represents one building under the considered electricity tariff, which is represented by the color scale}
\label{fig:ssvssc}
\end{figure}

Figure \ref{fig:ssvssc} presents the energy matching chart, with the self-consumption and self-sufficiency at the building level for each network. The points in the diagonal line represent net-zero buildings on yearly basis (i.e., buildings that produce as much energy as they consume), whereas points above the diagonal line are net producers and those below are net consumers on yearly basis. In the three networks analyzed and regardless of the electricity tariff used, most buildings are net producers. Moreover, as a result of the storage installed, tariffs with a capacity-based component increase self-sufficiency up to 45, 29, and 28 percentage points for the rural, semi-urban, and urban zones, respectively. 

\subsection{Tariff's impact on the grid}
\begin{figure}
\centering
\includegraphics[scale=0.4]{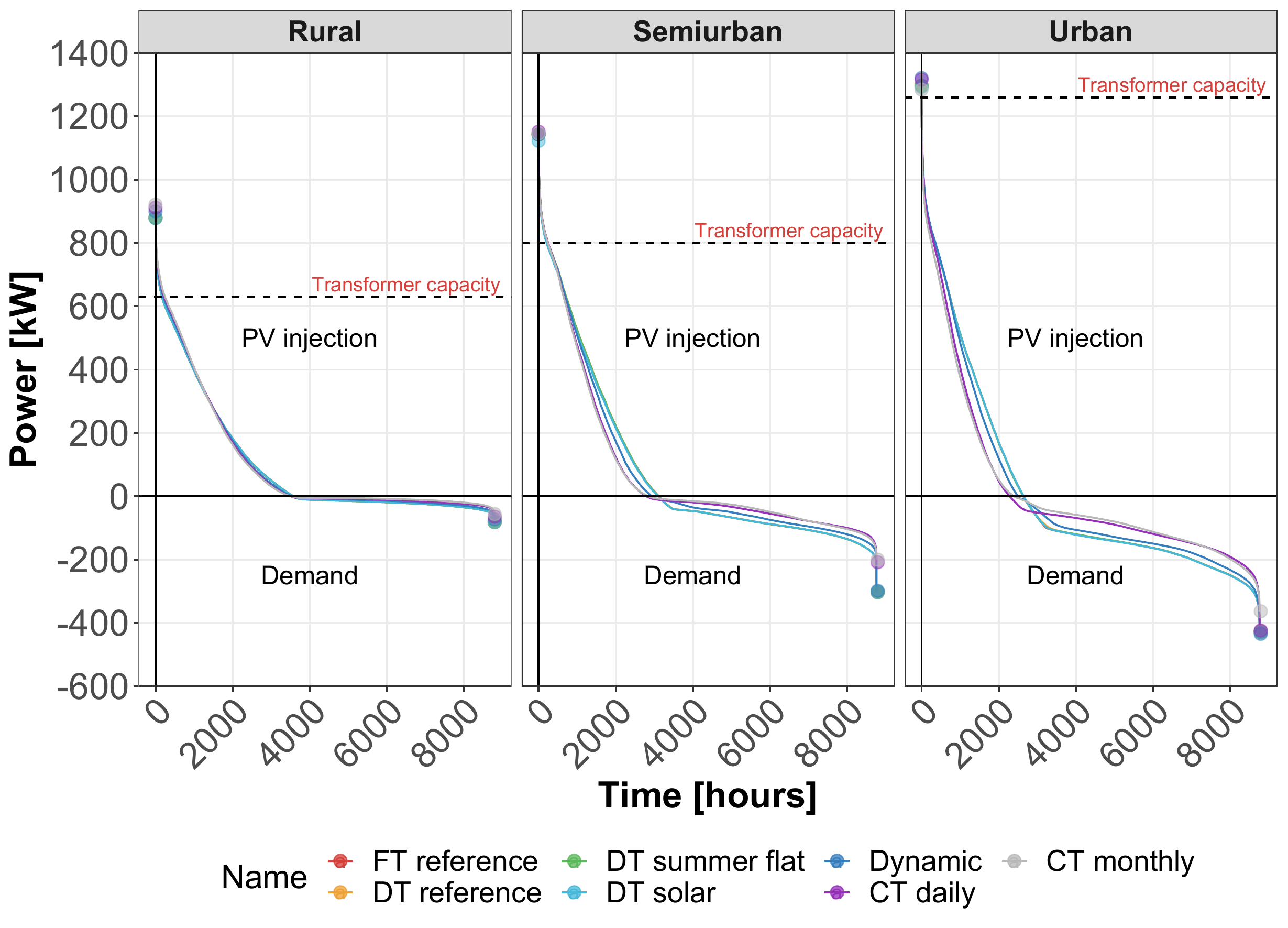}{\centering}
\caption{Load duration curve at the transformer. Dots on the vertical axis indicate the total installed PV capacity per scenario. Negative values indicate power flow from the high-voltage toward the low-voltage side}
\label{fig:loaddur}
\end{figure}

Figure \ref{fig:loaddur} presents the violations of the transformer power capacity, where values over the transformer capacity (indicated with dashed lines in the figure) indicate reverse power flow from the LV network towards the MV network. In the case of the rural zone, the maximum PV potential exceeds the capacity of the local transformer (630\,kVA), however, the transformer is overloaded only between 177 and 261 hours p.a. (depending on the electricity tariff). In the semi-urban zone, the situation is similar, and the transformer is overloaded between 223 and 260 hours p.a. In the case of the urban zone, however, the transformer is overloaded only for 1.75 hours. In general, the transformers of the three zones are overloaded less than 3\% of the hours per year. Finally, the tariff CT monthly is the tariff that reduces the most the peaks of demand. When the maximum peak demand results from the tariff CT monthly are compared to the FT reference, the peak is reduced from 81 to 56\,kW (i.e. 30.9\%) in the rural zone, in the case of the semi-urban zone the reduction is from 302.7 to 199\,kW (i.e. 34.2\%), and in the case of the urban network from 426 to 363\,kW (i.e. 14.8\%).

\begin{figure}
    \centering
    \includegraphics[scale=0.4]{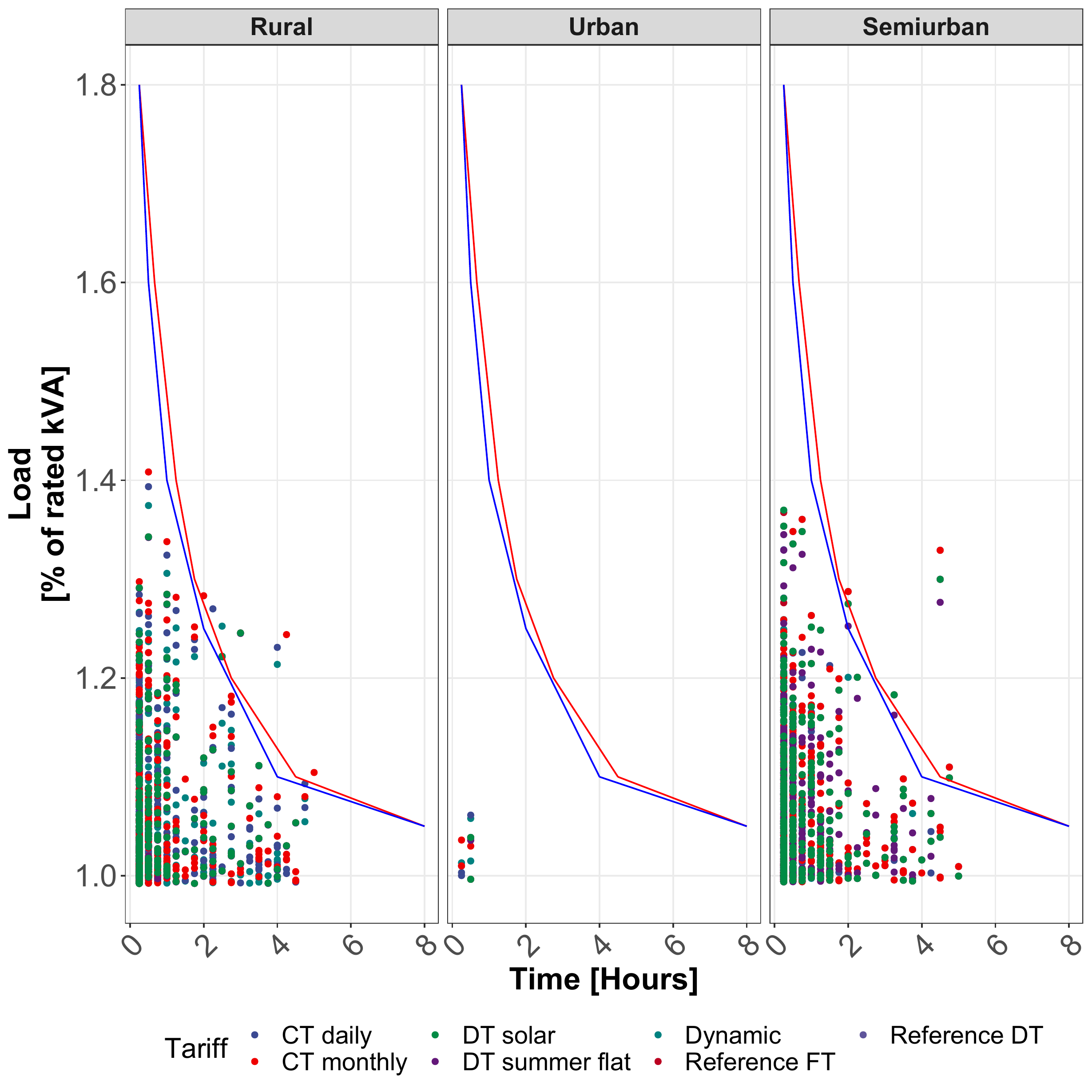}
    \caption{Permissible overload for varying periods of time, based on \cite{trafos2000} (in red) and \cite{RomandeEnergie_trafos} (in blue).}
    \label{fig:over}
\end{figure}

Transformers may be operated above average continuous temperatures, typically above 95 $^{\circ}$C for transformers rated at 55 $^{\circ}$C and 110 $^{\circ}$C for transformers rated at 65 $^{\circ}$C, for short times without affecting the normal life expectancy if most of the time they are operated at lower temperatures \cite{trafos2000}. Figure \ref{fig:over} presents the number of events where the transformer is overloaded (mainly due to PV injection) depending on the tariff applied. The permissible overload for different periods of time, where the transformer is operated without affecting the normal life expectancy, are below the red (based on \cite{trafos2000}) and blue (based on \cite{RomandeEnergie_trafos})  curves. The main hypotheses behind the curves are an ambient temperature of 30°C and an initial load factor of 80\%, which may be an extreme case (if the initial load factor is lower, the curve is displaced to the right). Based on the results, there are only a few moments throughout the year when the transformers of the rural and semi-urban networks are overloaded beyond the points of safe operation (i.e. affecting the normal life expectancy of the transformer). However, the provision of flexibility services to the DSO, PV curtailment, or capacity-based restrictions on the export could further alleviate the PV injection and admit a full deployment of PV within the two zones without the need for grid reinforcement. In the case of the urban network, no measure is needed since the transformer can manage the limited overloads without reducing its normal life expectancy.

\begin{figure}
\centering
\includegraphics[scale=0.5]{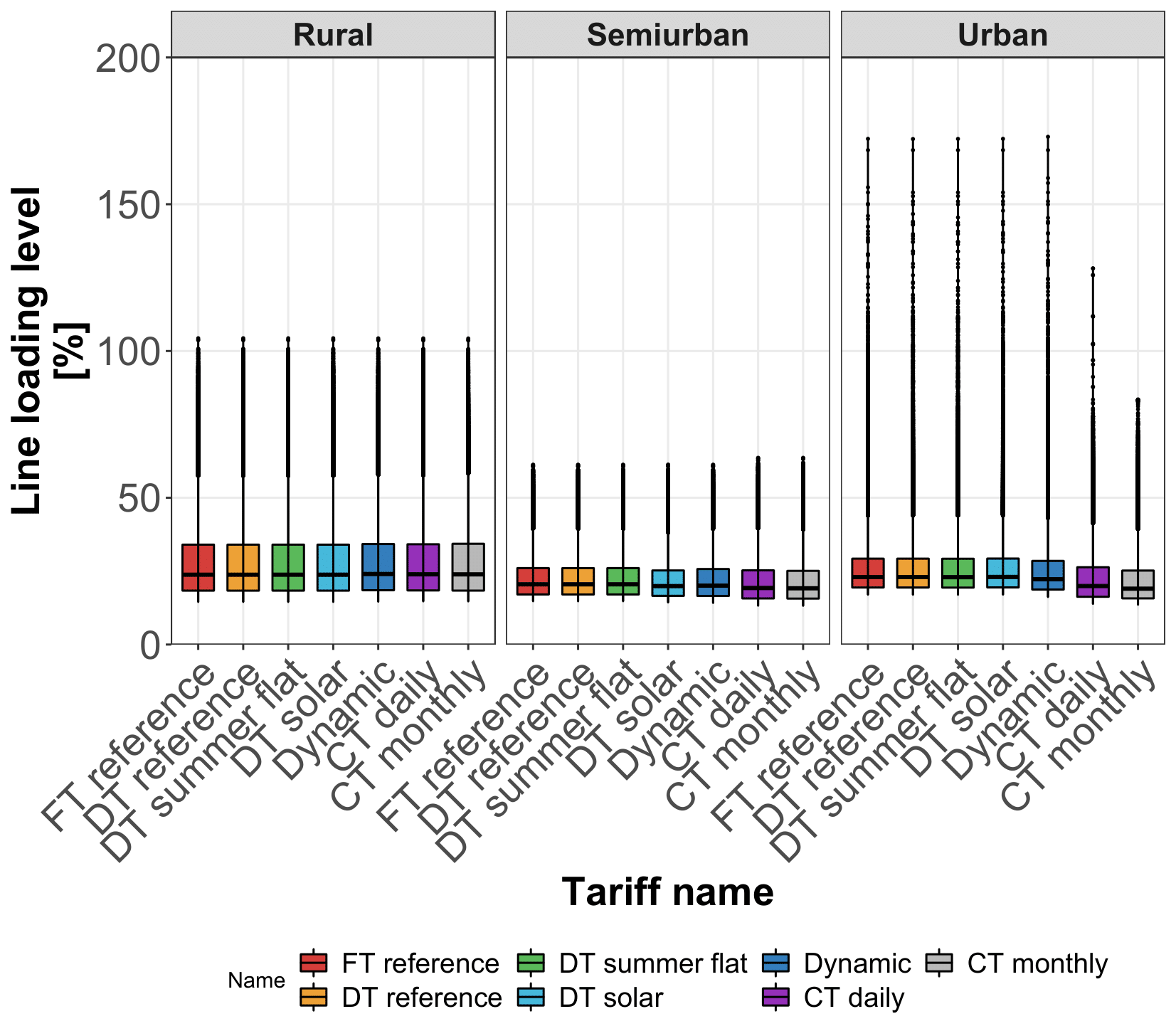}{\centering}
\caption{Line loading level distribution ($95^{th}$ percentile); Box plots show the median (horizontal line) and the IQR (box outline). The whiskers extend from the hinge to the highest and lowest values that are within 1.5 × IQR of the hinge, and the points represent the outliers.}

\label{fig:lineload}
\end{figure}

\begin{figure}
\centering
\includegraphics[scale=0.5]{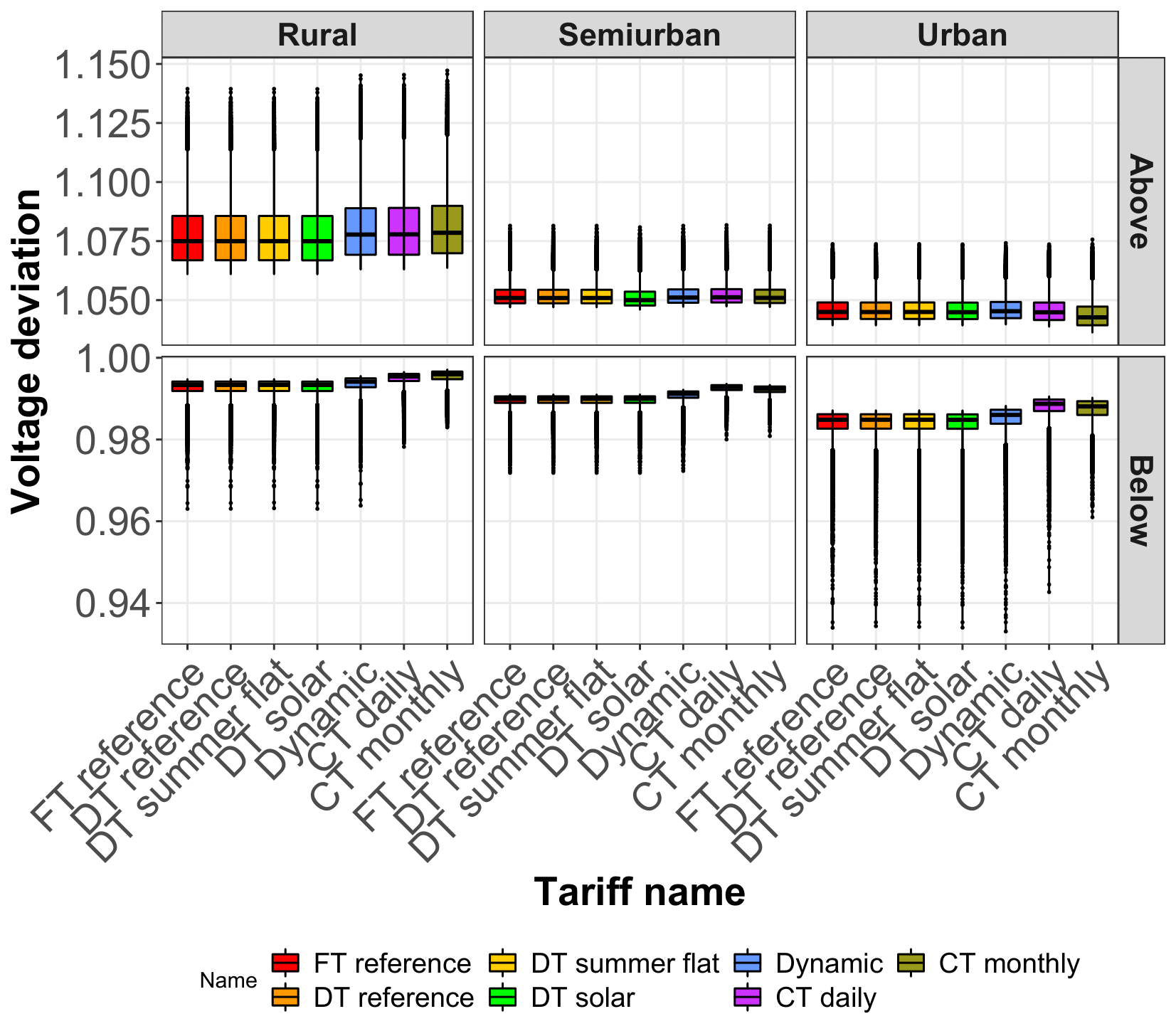}{\centering}
\caption{Voltage deviation distribution ($95^{th}$ percentile); Box plots show the median (horizontal line) and the IQR (box outline). The whiskers extend from the hinge to the highest and lowest values that are within 1.5 × IQR of the hinge, and the points represent the outliers.}
\label{fig:volt}
\end{figure}

The results of the current of the lines and voltage magnitude at the node level are presented in Figures \ref{fig:lineload} and \ref{fig:volt}. The rural network presents a median line loading level of 23.9\%, and only three data points are above the 100\% line loading level, regardless of the tariff used. For the semi-urban zone, the maximum line loading level is 64\%, with a median value of 20\%. In the urban zone, the median line loading level reaches 22\% but the maximum line loading level is 173\% for all tariffs except for the two references FT and DT, which reach 172\% and 83\%, respectively. The line overloading occurs mainly during the winter months (primarily, November, December, and February), and affects ten lines (out of 93). It is therefore not related with PV excess injection. 

Positive voltage deviations are critical in the rural network, where there are up to 18420 data points above the (rather soft) limit of 1.1\,p.u., violating current European and Swiss regulations \cite{hartvigsson2021estimating}. In the case of semi-urban and urban networks, there are no voltage violations due to PV injection. No evidence of voltage violations due to demand in the studied zones was found. However, the use of the dynamic tariff and tariffs with a capacity-based component present statistically significant differences with the reference tariffs (p-value $\leq$ 2e-16), resulting in median values for voltage deviation closer to 1, meaning that the use of such tariffs most probably will result in lower voltage deviations.

\subsection{Tariff's impact on the grid cost recovery}
Table \ref{tab:gridcost} presents the grid cost recovery per network. For the aggregated results the volumetric tariffs (i.e., summer flat, solar, and dynamic) are slightly above the reference DT, surpassing the total grid cost recovery by 0.9\%, 0.3\%, and 0.3\%, respectively. In the cases of the two tariffs with a capacity-based component (i.e. CT monthly, and CT daily), the grid cost recovery falls behind the level of the DT reference by 5 and 1\%, respectively. These values can be further improved to match the required grid cost recovery levels, nevertheless, what is worth mentioning is that with a CT monthly of 16.4\,€/kW and a CT daily between 0.53 and 1.33\,€/kW, the DSO can recover the grid costs.

\begin{figure}
\centering
\includegraphics[scale=0.4]{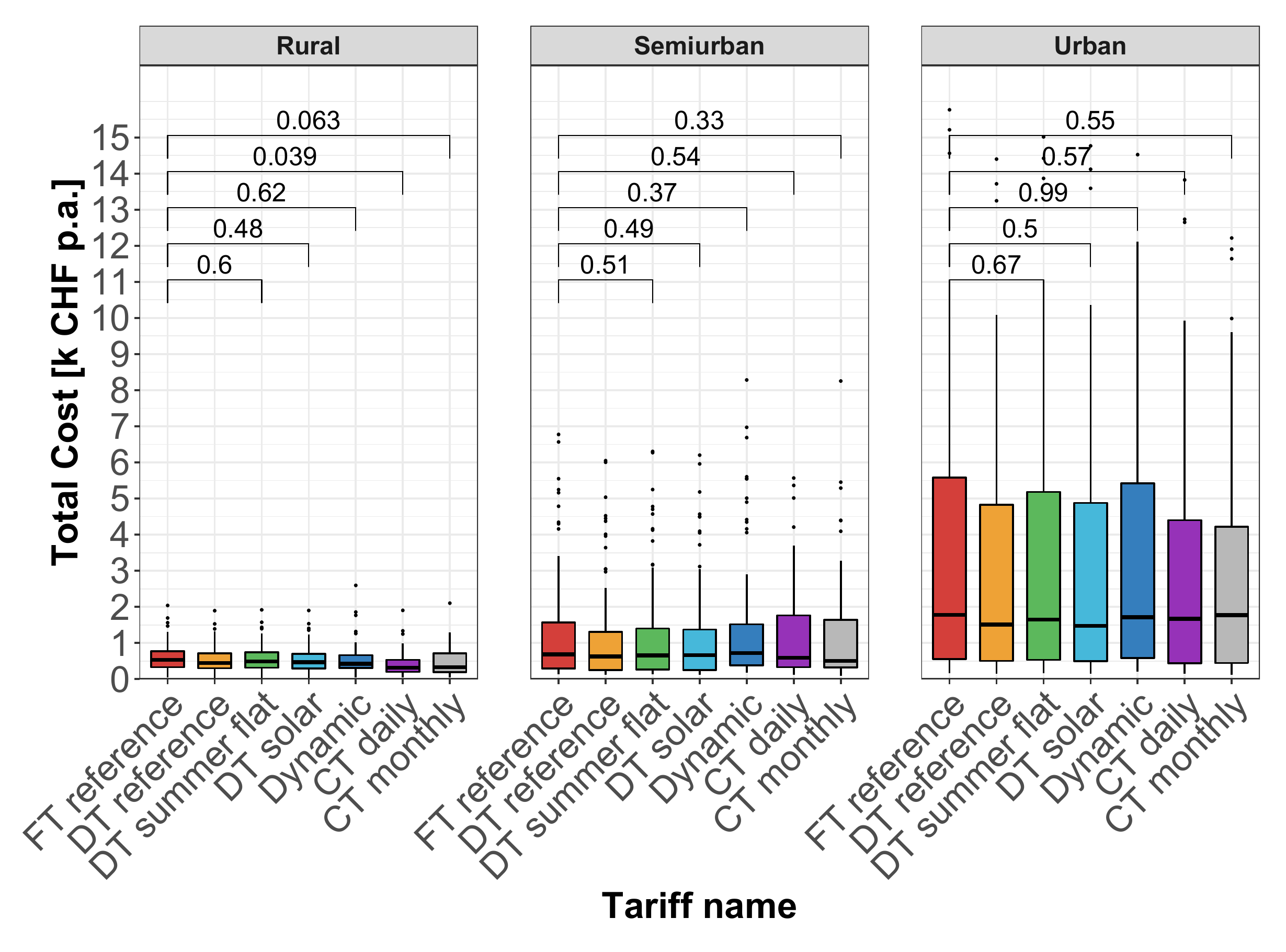}{\centering}
\caption{Box plot of the annual cost for each building depending on the network; Box plots show the median (horizontal line) and the IQR (box outline). The whiskers extend from the hinge to the highest and lowest values that are within 1.5 × IQR of the hinge, and the points represent the outliers. The individual p-values of the two-sided Wilcoxon tests with the Holm procedure to control the family-wise error rate are reported in the figure.}
\label{fig:ind}
\end{figure}

The total bill cost per building per year is presented in Figure \ref{fig:ind}. The total bill cost per building does not present statistically significant differences even when tariffs with a capacity-based component are used (p-values$>$0.1 when compared to the two references). This result indicates that when changing the reference tariff for another one of the five proposed in this study, the total bill cost per building remains similar. The only exception is the rural LV network under a CT daily when compared to the FT reference, where there is a significant difference (p-value=0.039), and therefore, a lower bill cost should be expected. Nevertheless, when the same comparison is done regardless of the network, there is no statistically significant difference between the two cases, which indicates that at the aggregated level there are no differences in the total bill per building. We acknowledge that the absence of statistically significant differences among the studied tariffs is true under a scenario with high penetration of PV and storage. 

\begin{table}[ht]
\centering
\scalebox{0.8}{
\begin{tabular}{lrrrrrrr}
  \hline
Network & FT reference & DT reference & DT summer flat & DT solar & Dynamic & CT daily & CT monthly \\ 
  \hline

Rural & 9093 & 10581 & 10586 & 10606 & 8957 & 8811 & 10948 \\ 
  Semiurban & 43979 & 51005 & 51174 & 51358 & 50704 & 48945 & 53106 \\ 
  Urban & 94049 & 109411 & 109195 & 110543 & 111932 & 104523 & 105333 \\  \hline 
  Total &  147121 &	170997 &	170955 &	172507	& 171593 &	162279 &	169387\\
   \hline
\end{tabular}}

\caption{Grid cost recovery per network and tariff, in € per year.}
\label{tab:gridcost}
\end{table}

\section{Discussion}\label{discussion}
Our study, bridging real LV network data with robust energy modeling, indicates that alternative tariffs can drive the adoption of storage, and recover the grid costs without causing relevant economic differences for the customers in LV networks. However, they only marginally mitigate the impact of PV on the network. Our findings have four key implications for DSOs, policy-makers, and academics.

First, we found that all the assessed tariffs in this study perform similarly or better than the two references in terms of driving the adoption of PV and storage. In the case of the two more traditional volumetric tariffs (DT summer flat and DT solar), they do not provide further incentives to increase the amount of PV or battery capacity within the LV networks. This result calls for complementary incentives to increase the share of storage such as financial rebates if the electricity tariff structure cannot be drastically modified for economic or social reasons. On the contrary, the Dynamic tariff, based on the expected aggregated load curve of the DSO, drives the adoption of additional capacity of PV and storage in the three LV networks, helping to reduce the energy imports but only marginally reducing the energy exports. Finally, the tariffs with a capacity-based component are the tariffs that help the most to increase the installation of PV and storage, being particularly important the effect of the tariff with a billing horizon of one month. Moreover, under this same tariff (i.e. CT monthly), in the three LV networks, the energy imports and exports are reduced the most, with respect to the reference tariffs. In terms of self-consumption and self-sufficiency, most of the buildings are net producers across the three LV networks, which points towards a reduction in energy needs in the future. The tariffs with a capacity-based component were found to increase both indicators when compared to the references, mainly due to the higher penetration of batteries.

Second, we found that in general, the assessed tariffs display a similar impact of PV on the LV network in terms of PV hosting capacity, i.e. the maximum amount of PV that can be accommodated in a power grid within current regulations without requiring reinforcement. Regardless of the tariff, the three analyzed LV networks are reaching the limits of their PV hosting capacity, the rural and semi-urban networks are the networks with higher overload. However, the three analyzed transformers would mostly operate within the permissible zones, without affecting their normal life expectancy and therefore without the need for reinforcement. Despite the PV penetration for the three zones being over 90\% and therefore close to the technical maximum, the theoretical maximum PV penetration used in this study only takes into account the maximum rooftop potential, neglecting further facade PV additions that could then contribute to exceeding the networks' PV hosting capacity. In such a case, the provision of flexibility services to the DSO, PV curtailment, or capacity-based restrictions on the export could alleviate further the PV injection and admit a higher deployment of PV within the three zones without the need for grid reinforcement. The use of tariffs with a capacity-based component was found to result in lower voltage deviations above 1\,p.u., when compared to the volumetric tariffs, in the rural setup. These results, are in line with the findings of Hartvigsson \textit{et al.} \cite{hartvigsson2021estimating}, for Sweden, Germany, and the UK, where voltage increase is the main factor limiting the PV hosting capacity. This result is more noticeable when a more stringent limit for the voltage limits (e.g. a 5\% deviation) is used since for most of the $95^th$ percentile voltage distribution, it is violated in all three networks when PV electricity is exported and in the Urban setting when electricity is drawn from the grid. Similarly, we found that most problems stem from the rural networks, where lines are longer, the available area for PV is higher and the demand is lower than in the urban networks \cite{gupta2020spatial, hartvigsson2021estimating}. 

Third, the mixed tariffs with volumetric and capacity-based components allow the DSOs to recover the grid cost, without incurring in statistically relevant differences for the customer in a scenario with high PV penetration and with some storage. However, we acknowledge that socio-economical differences, as highlighted by Azarova \textit{et al.} \cite{azarova2018exploring}, should be taken into account at the moment of proposing a change of tariffs to the clients. Moreover, the tariffs assessed in this study concern only the demand side, since as explained by Bloch \textit{et al.} \cite{bloch2019impact}, a capacity-based tariff for both, import and export does not encourage investment in battery systems, since the cost is dominated by the PV injection power, which can be simply curtailed. Additionally, this tariff would charge twice the user for the use of the grid, once when consuming electricity from the grid, and the second when injecting PV electricity into the grid.

Fourth, in the Swiss context, rooftop PV plays a paramount role in the energy transition. Only a reduced amount of land is available for solar farms due to scarcity, citizen opposition, and/or law mandates. Our results highlight the problems that DSO may encounter, mainly in rural areas. This study emphasizes the need to act on the export side to help to mitigate the impact of PV.

While we have considered as reference cases three low-voltage networks representative of Switzerland, further case studies including other regions and countries could lead to conclusions that could be generalized. We considered an optimization following an economic rational behavior for all buildings and considered PV and battery installations to be done at once, whereas in reality, most probably not every building will be able to install PV on the rooftop either for economic or non-economic reasons hindering the national objectives towards PV deployment. Moreover, this process is done progressively leaving some time for DSOs to adjust their decisions in terms of tariff design and low-voltage system operation. We assumed perfect foresight of the demand; however, the longer the period over which the peak demand is observed, the higher the incertitude of the peak demand would be. Our research can also be a starting point for future research considering the grid-friendly operation of low-voltage networks considering other distributed energy resources such as EVs and HPs that could help to increase the PV hosting capacity. Moreover, since most of the problems arise from the export side, the exploration of capacity tariffs to the export or variable export prices could give a better idea of how to mitigate such problems at the distribution level. Finally, curtailment or curtailment incitation (e.g., higher export tariffs if the DSO can control PV curtailment or pays for curtailed energy) could lead to more cost-efficient results and have higher social acceptance than capacity-based tariffs.


\section{Conclusions}\label{conclusions}
In this study, we did a comprehensive analysis that integrates the effects of five tariffs on the optimum size and operation of PV and battery systems, as well as on different typical LV networks, namely, rural, semi-urban and urban in Switzerland. The considered tariffs include two traditional (volumetric) double tariffs that seek to promote PV self-consumption (DT solar and DT summer flat), as well as a volumetric dynamic tariff proportional to the aggregated expected load curve of the DSO for 2025. Moreover, we analyze two mixed tariffs including a volumetric and a capacity-based component (in \,€/kW), with two billing horizons, one monthly (CT monthly) and one daily (CT daily). These five tariffs are compared to two reference tariffs (flat and double tariff) that are currently used by the DSO.

The results show that traditional volumetric tariffs perform similarly to the two reference tariffs (flat and double tariff) without adding any benefit, in terms of the adoption and operation of distributed PV-coupled battery systems. Only alternative tariffs, such as dynamic tariffs (based on the aggregated expected load of the DSO, in this study) and tariffs with a capacity-based component, were found to induce a higher adoption of PV and storage.

In line with the literature, we found the major issues to arise mainly in rural grids, and to involve voltage violations and reverse power flow. Additionally, the maximum amount of PV that can be accommodated in a power grid without requiring reinforcement was found to be higher than the maximum capacity of the transformer, given that the overloading events of the transformer are short in timespan. In this sense, and since the main problems for the distribution grid, in general, stem from PV injection, the electrification of the heating and transport sector may relieve such problems if additional demand occurs during surplus times. Otherwise, solutions involving the management of the PV export should be considered. For instance, considering PV curtailment, capacity-based tariffs to the contracted capacity, capacity-based tariffs to the export, or local flexibility to mitigate the crucial hours of the year and in this way help to relieve transformer overloading.
\section{Acknowledgements}
This research project was financially supported by Romande Energy. The funding source had no involvement in the preparation of the article, in the study design, the collection, analysis, and interpretation of data, nor in the writing of the manuscript. Furthermore, we thank Arnoud Bifrare, Stéphanie Weiler, and Maret Julien for their insights and comments on this study.
\section{Declaration of competing interest}
The authors declare that they have no known competing financial interests or personal relationships that could have appeared to influence the work reported in this paper. 
\section{CRediT authorship contribution statement}
\textbf{Alejandro Pena-Bello:} Conceptualization, Data curation, Methodology, Validation, Investigation, Formal analysis, Writing - original draft, Visualization. \textbf{Robin Junod:} Software and Investigation. \textbf{Christophe  Ballif:} Writing - review \& editing, Funding acquisition. \textbf{Nicolas Wyrsch:} Conceptualization, Supervision, Writing - review \& editing, Funding acquisition, Project administration. 
\section{Data Availability}
The model will be made available on reasonable request. The LV network data is subject to a confidentiality agreement.

\bibliographystyle{vancouver}
\bibliography{biblio}

\end{document}